\documentclass[12pt,onecolumn,aps,pra,floatfix,superscriptaddress,preprint, showpacs,groupedaddress]{revtex4-2}

\usepackage{upgreek}
\usepackage{graphicx}
\usepackage{graphics}
\usepackage{amssymb}
\usepackage{amsmath}
\usepackage{epsfig}
\usepackage{latexsym}
\usepackage{color}
\usepackage{rotating}
\usepackage{subfigure}
\usepackage{float}
\usepackage[breaklinks=true]{hyperref}
\usepackage[labelfont=bf]{caption}
\usepackage{ragged2e}
\DeclareCaptionJustification{justified}{\justifying}
\hypersetup{
  colorlinks   = true, 
  urlcolor     = blue, 
  linkcolor    = blue, 
  citecolor   =  red 
}

\usepackage{soul} 

\begin{document}

\title{Dynamic interaction between chiral currents and surface waves in topological superfluids: a pathway to detect Majorana fermions?}

\author{S. Forstner$^{1,2}$\footnote{stefan.forstner@icfo.eu}}

\author{H. Choi$^{3,4}$\footnote{h.choi@kaist.ac.kr} 
}
\author{G.I. Harris$^{1}$ 
}
\author{A. Sawadsky$^{1}$ 
}
\author{W.P. Bowen$^{1}$ 
}
\author{C.G. Baker$^{1}$\footnote{c.baker3@uq.edu.au}}

\affiliation{$^1$ARC Centre for Engineered Quantum Systems, School of Mathematics and Physics, The University of Queensland, St Lucia, QLD 4072, Australia}
\affiliation{$^2$ICFO-Institut de Ciencies Fotoniques, The Barcelona Institute of
Science and Technology, Castelldefels (Barcelona) 08860, Spain}
\affiliation{$^3$Department of Physics, KAIST, Daejeon 34141, South Korea}
\affiliation{$^4$Graduate School of Quantum Science and Technology, KAIST, Daejeon 34141, South Korea}

\begin{abstract}

Despite extensive experimental efforts over the past two decades, the quest for Majorana fermions in superconductors remains inconclusive. We propose an experimental method that can conclusively confirm, or rule out, the existence of these quasiparticles: Firstly, we shift focus from \mbox{superconductors}, whose very topological nature is disputed, to the unambiguous topological superfluid $^3$He. Secondly, we identify the interaction between surface waves and the  
chiral Majorana current in the bulk of a topological superfluid of varying density.
The proposed experiment provides a path towards the detection of the Majorana fermion, an 80-year-old theoretical prediction. It is realistically achievable based on the advent of microscopic superfluid resonators coupled to optical cavities.
The proposal may open the door to experiments ranging from simulations of exotic cosmological particles to topological acoustics and fault tolerant quantum computing. 
\end{abstract}

\maketitle

\section*{Introduction}
One of the major breakthroughs in modern condensed matter physics is the incorporation of topology in material classifications, beyond traditional classifications based on symmetry~\cite{chiu_classification_2016}. However, despite the successful experimental realization of topological insulators~\cite{Bernevig_QSHE_HgTe_Theory_2006,Konig_QSHE_HgTe_Exp_2007} and subsequently topological semimetals~\cite{Burkov_TopoSemimetal_2016,Liu_topologicalDirac_2014,Xu_topologicalWeyl_2015}, topological superconductors remain an elusive goal~\cite{Qi_Topo_Ins_Sup_RMP_2011,Sato_TopoSuper_2017}.
Perhaps the most intriguing prediction about topological superconductors is that their boundaries are a potential host to a quasiparticle known as the Majorana fermion~\cite{Qi_Topo_Ins_Sup_RMP_2011,Sato_TopoSuper_2017}.
The Majorana fermion, a particle with the unusual property of being its own antiparticle, was first theoretically proposed more than 80 years ago \cite{Majorana_Teoria_1937}. Despite the lack of experimental observation, its theoretical properties have been intensely studied, as have their implications 
for particle physics~\cite{Makinen_vortices_2018,Volovik_universe_2009} 
and fault-tolerant quantum computing~\cite{Nayak_Anyons_2008,Biao_chiral_2018,Ivanov_nonabelian_2001,DasSarma_ZeroModes_2015}.
Direct synthesis of such a superconductor~\cite{Mackenzie_Sr2RuO4_RMP_2003} has been found to be 
unsuccessful \cite{Pustogow_Sr2RuO4_OP_2019}.
A more subtle strategy of combining two or more materials \cite{Fu_Kane_TSC=SC+TI_2008}, each exhibiting non-trivial topology and superconductivity separately, led to observations that suggested the existence of Majorana fermions \cite{Mourik_Majo_Ferm_Hybrid_2012,NadjPerge_MF_FM+SC_2014,He_chiral_MF_2017,Lutchyn_MZM_SC_Semi_2018}. However, the interpretation of these observations has since been questioned \cite{Kayyalha_Absence_2020, Zhang_Retraction_2021, Yu_NonMZM_2021,Valentini_Majoranalike_2022}.

While nearly all of these efforts have been focussed on topological superconductors
\cite{Read_Green_FQHE_2000,Nayak_Anyons_2008}, the advent of topology also shed new light on a long-known system: Superfluid helium-3.
Unlike in any known superconductor, the $p$-wave nature of superfluid $^3$He's order parameter is unequivocally established in its $A$- and $B$-phases. Thus, it is widely expected to contain Majorana fermions.
There are a number of experimental reports of the existence of surface states in superfluid $^3$He \cite{Aoki_SABS_TAI_Measurements_2005,Choi_Surface_Specific_Heat_2006,ikegami_chiral_2013} which suggest the existence of the Majorana fermions. However, direct experimental confirmation of the topological nature of the bulk order parameter or the Majorana nature of these boundary states is still an outstanding challenge.
One of the most powerful ways in which non-trivial topology reveals itself in a condensed matter system is in the form of a quantized conductance whose quantization is strongly tied to the 
topological invariants of a system. Given the analogy between the quantum Hall system and topological chiral superconductivity, quantized Hall conductivity is expected with the chiral edge current carried by Majorana fermions. However, quantum Hall measurements rely on the presence of charge in the system and are thus inapplicable to liquid $^3$He.

In this work, we propose an experiment that can directly and unambiguously measure the chiral Majorana current by leveraging the to-date barely considered spatially distributed current, which arises in response to local density variations in a $p$-wave superconductor or superfluid. We show how this effect arises dynamically due to sound modes in superfluid \mbox{$^3$He-$A$}, and how it can be directly detected via its dispersive interaction with these very sound modes \cite{schechter_observation_1998}. 
In a two-dimensional picture, sound-induced deformations in the surface can be thought of as fluctuations in planar density. Such fluctuations induce a transverse current in the direction perpendicular to the density gradient, an effect reminiscent of the transverse current induced in response to a voltage gradient in a Hall system. We term this spatially- and temporally varying effect the \textit{dynamic} chiral current, as opposed to the conventionally predicted \textit{static} chiral edge current at the superfluid boundary \cite{byun_measuring_2018}. 
This dynamical effect can be interpreted as a manifestation of the dissipationless quantum Hall viscosity \cite{Avron_QuantumViscosity_1995}  with the strain rate and subsequent momentum current being orthogonal to each other. Despite significant research efforts, this effect has been obscured in electronic quantum hall systems by the presence of the lattice \cite{Barkeshli_DisspationlessHall_2012,Ramamurthi_Topological_Semimetals_2015,Rao_HallViscosity_2020}.
We show that this transverse current
couples orthogonal pairs of superfluid sound modes and as a result lifts the degeneracy between them. 
We calculate the magnitude of this effect and investigate potential obstacles to unambiguous experimental detection. We find that its magnitude scales inversely with the resonator size, so that to obtain a clear signature, microscale superfluid resonators are essential. Recent advances in miniaturized superfluid thin-film resonators, in which superfluid surface modes are read out optically~\cite{harris_laser_2016,mcauslan_microphotonic_2016,sachkou_coherent_2019,he_strong_2020,sawadsky2022engineered}, hence open the door to detecting this deformation-induced Majorana current.

Experimental observation of the dynamical chiral current in superfluid $^3$He would provide, for the first time, incontrovertible evidence of quantized Hall effect and chiral Majorana fermions. Moreover, the proposed experiment would constitute the first realistic method to study dissipationless quantum Hall viscosity. This could open the door to employing these elusive quasi-particles to study non-Abelian statistics, to simulate exotic fundamental particles  and cosmological phenomena~\cite{Makinen_vortices_2018,Volovik_universe_2009}, to realize topological acoustics~\cite{yang_topological_2015,souslov_topological_2017}, and may even enable fault-tolerant topological quantum computing~\cite{Nayak_Anyons_2008,Biao_chiral_2018,Ivanov_nonabelian_2001,DasSarma_ZeroModes_2015}.
\section*{Results}
\subsection*{Coupling between chiral edge current and sound modes}

The mass current density in superfluid $^3$He-$A$ at zero temperature is given by \cite{cross1975generalized,mermin_cooper_1980}:
\begin{equation}
\vec{g}=\rho\,\vec{v}+\frac{\hbar}{4\,M}\vec{\nabla}\times \rho \,\vec{l}-\frac{\hbar}{2\,M}\,c_0\, \vec{l}\left(\vec{l}\cdot \vec{\nabla}\times\vec{l}\,\right)
\label{Eqmasscurrentmermin}
\end{equation}
where $\vec{v}$ is the superfluid velocity, $\hbar$ the reduced Planck constant, $M=5.0\cdot10^{-27}$~kg the mass of a helium-3  atom, $\rho$ the local superfluid density, $\vec l$ the local anisotropy axis, and $c_0$ 
is an integral over single-particle momentum states which asymptotes to $\rho$ as the temperature goes to zero~\cite{Combescot_Supercurrent_3He-A_1983} . In the case of a helium thin film, $\rho$ can be understood as the bulk density $\rho_0$ of superfluid helium-3, locally scaled with the film thickness variation: $\rho=\rho_0\cdot(1+\eta/h_0)$, with $\rho_0=82$~kg/m$^3$, $h_0$ the equilibrium film height, and $\eta$ the film height perturbation due to the acoustic wave. The first term corresponds to an irrotational mass current, such as that arising from center-of-mass motion or acoustic waves. The second term describes the chiral current, in analogy to the bound current density $j$ in a conductor of varying magnetization $\vec{M}$, ($j\propto |\vec{\nabla}\times \vec{M}|)$, and can be thought of as arising due to the Cooper pairs' orbital angular momentum \cite{mermin_cooper_1980}.  The third term, which generates an essentially out-of-plane current, is neglected in the following, as its coupling to the in-plane flow due to sound modes is minimal.

In this work, we shall consider the case of a helium-3 surface-sound resonator and focus on the role of the second term in Eq.~(\ref{Eqmasscurrentmermin}). Spatial variations in $\vec{l}$ or in $\rho$ lead to net mass currents.
The direction of the current depends on the preferred orientation of the orbital angular momentum $\hat{l}$, which is 
to first order either aligned or anti-aligned with the resonator's surface normal $\hat{z}$~\cite{ambegaokar_landau-ginsburg_1974}.
Here, we consider the large wavelength, small amplitude limit, where the acoustic wavelength $\lambda$ is much larger than the film thickness $h$, which is in turn much larger than the out-of-plane displacement amplitude $\eta$ of the film. This implies that the motion of the fluid is essentially in-plane and hence out-of-plane motion is negligible (see Methods 
for details). Such a condition arises naturally in experiments, where a resonator substrate is coated with a film of superfluid helium (see e.g. \cite{ellis_quantum_1993,sachkou_coherent_2019}). 

\begin{figure}
\centering
\includegraphics{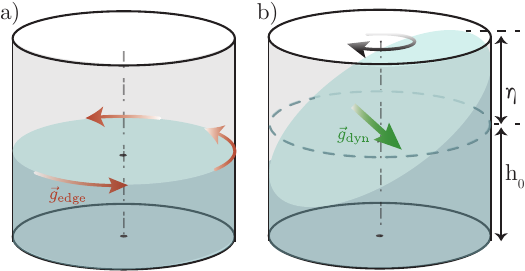}
\caption{Illustration of static and dynamic chiral currents. (a) Static chiral edge current, arising only at the boundary of the resonator. Red arrows show direction and location of the induced flow, black arrow on top shows the rotation direction of the fluid antinode.
(b) Dynamic chiral current, arising as a result of the acoustic mode-induced height variation, and present throughout the resonator. The green arrow shows the direction of the induced flow, and the black curved arrow shows the rotational direction of the clockwise rotating sound mode. $h_0$ and $\eta$(not to scale) indicate equilibrium film height and acoustic perturbation, respectively.}
\label{Figure1}
\end{figure}

Under the assumptions of the previous paragraph, the second term in Eq.~(\ref{Eqmasscurrentmermin}) can be expanded as
\begin{equation}
\frac{\hbar}{4\,M}\vec{\nabla}\times \rho \,\vec{l} = \vec{g}_\mathrm{edge} + \vec{g}_\mathrm{dyn}.
\label{EQWarwick}
\end{equation}
Here, $\vec{g}_\mathrm{edge}$ is the conventional static chiral edge current 
\cite{byun_measuring_2018,sauls_surface_2011}, while $ \vec{g}_\mathrm{dyn}$ is a dynamic current which arises due to the presence of the sound modes and which we evaluate for the first time in this work. 
These two contributions to the in-plane mass current are  illustrated in Fig. \ref{Figure1}.
The static mass current, shown in Fig. \ref{Figure1} (a), results from the steady-state variations in the film's thickness due to physical confinement within the resonator geometry. In the case where the film thickness and density are uniform, and the thickness drops to zero at an abrupt specular boundary, 
it corresponds to the ground-state angular momentum $N \hbar/2$ \cite{stone_edge_2004, sauls_surface_2011}, with $N$ the number of helium nuclei contained in the film.
The dynamic mass current, shown in Fig. \ref{Figure1} (b), is proportional to the gradient of the (temporally and spatially varying) superfluid film height due to the sound mode. 

Both the static and the dynamic contributions to the chiral edge current couple to acoustic waves via their direct overlap with the flow fields of the sound modes. In the static case, the chiral mass current leads to a time-independent flow field that superposes with the flow field of acoustic waves. The energy of an acoustic mode increases if the interference is constructive, and decreases if it is destructive. On the other hand, the dynamic chiral current arising from the density perturbation due to one acoustic mode can interfere with the flow field of another mode, leading to a ``dynamic coupling" between acoustic modes.

\subsection*{Dynamic coupling}
\label{subsec:Dynamic}

We investigate first the effect of the \emph{dynamic} coupling in the analytically tractable case of a circular superfluid surface-wave resonator \cite{atkins_third_1959}. The restoring forces are surface tension, the van-der-Waals force, and gravity, either of which can be dominant depending on resonator radius, film thickness, and reservoir height. 
Note that while we consider surface waves here, this discussion is also valid for density waves, such as first sound, with the appropriate transformations (see Supplementary Note 1). 
For a thin-film superfluid surface-wave resonator with equilibrium film height $h_0$, using simple vector calculus (see Methods),
the dynamic part of the chiral current can be written as (c.f. Eq. \eqref{EQWarwick})
\begin{equation}
    \vec{g}_\mathrm{dyn}=-\rho_0\frac{\hbar}{2\,M\,h_0}\hat{z}\times \vec{\nabla}\eta+g_\mathrm{torque},
    \label{eqinplanemasscurrent}
\end{equation}
where $\eta\ll h_0$ is the surface-deflection due to the sound wave (see Fig. \ref{Figure1}), and $g_\mathrm{torque}$ is the out-of-plane component of the chiral mass current, which is negligible for the resonators considered here (see Methods).
Without loss of generality, we have assumed the anisotropy axis $\vec{l}$ to be aligned with the normal to the surface $\hat{z}$, i.e. $\hat{l}= \hat{z}$.

\begin{figure}
\centering
\includegraphics{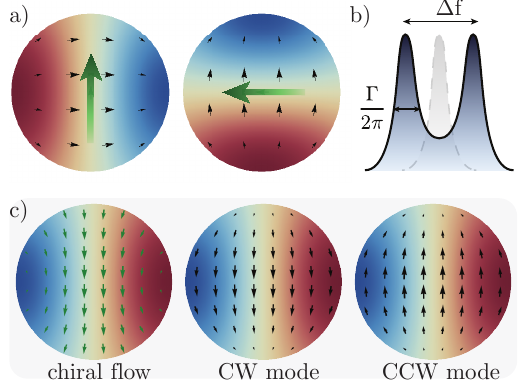} 
\caption{
Dynamic coupling mechanism. (a) Standing wave basis illustration of the lowest frequency ($m=1$;$n=1$) sound eigenmode of a circular resonator with free boundary conditions \cite{baker2016theoretical}. Background color: height profile $\eta\left( r, \theta \right)$; Black arrows: superfluid surface-sound flow field $\vec{v}_\mathrm{s}$; 
Green arrow: dynamically induced chiral flow $\vec{v}_{\mathrm{dyn}}$ due to Helium-3A's orbital angular momentum (see Eq. \ref{Eqmasscurrentmermin}). 
(b) Sound mode frequency splitting $\Delta f$ arising from Helium-3A's ground state orbital angular momentum. 
(c) Computed flow fields and displacement profiles for ($m=1$;$n=1$) Bessel mode with free boundary conditions (see also \cite{baker2016theoretical}). Left panel: Mass current vector field $\propto \vec{v}_{\mathrm{dyn}}$ obtained from Eq. \ref{Eqmasscurrentmermin}) (green arrows) caused by the height profile $\eta$ (displayed as the background color code). As expected, the  chiral flow 
is orthogonal to the height gradient. Middle \& right panels: Instantaneous flow fields for the clockwise and counter-clockwise sound modes respectively (black arrows). 
Note the good overlap between the chiral and  CW flow fields, and the destructive interference between the chiral and CCW flow fields.
}
\label{Figure2}
\end{figure}

To illustrate the effect of the chiral current on the acoustic eigenmodes, let us first consider the lowest frequency excitation of a circular resonator of radius $R$ with free boundary conditions, where the fluid oscillates back and forth from one side of the resonator to the other, corresponding to the lowest-frequency Bessel-mode  \cite{baker2016theoretical}. In the absence of chiral currents, this type of excitation can be described in the standing wave basis formed by the two frequency-degenerate normal modes shown in Fig. \ref{Figure2}(a), where the sound velocity field $\vec{v}_s$ is represented by the black arrows. The presence of the sound-induced density gradient $\eta$ leads via Eq. (\ref{eqinplanemasscurrent}) to an additional mass current $\vec{g}_{\mathrm{dyn}}$ (and associated flow field $\vec{v}_{\mathrm{dyn}}$) in a direction orthogonal to the height gradient, represented by the green arrows. This net mass flow will either interact constructively or destructively with the sound-induced flow, leading to a coupling between previously orthogonal standing wave modes. In the presence of this coupling, clockwise (CW) and counterclockwise (CCW) propagating surface-sound waves become  
the diagonal
mode basis.  The coupling introduces a frequency splitting between these waves, in a mechanism which is reminiscent of the third-sound splitting due to the presence of quantized vortices \cite{rudnick_observation_1969,ellis_observation_1989,ellis_quantum_1993,sachkou_coherent_2019,forstner2019modelling}. We propose that this splitting provides a mechanism by which the chiral current can be detected experimentally, e.g. by optomechanical readout \cite{sachkou_coherent_2019}.

This lifting of the degeneracy between normal modes is experimentally resolvable if its rate $\Delta f$ exceeds the acoustic mode's linewidth $\Gamma/(2\pi)$ \cite{forstner2019modelling, ellis_observation_1989, sachkou_coherent_2019},  as shown in Fig. \ref{Figure2}(b). A better understanding of the newly formed eigenmodes can be gained by shifting to the more appropriate CW and CCW rotating-mode basis (see Fig. \ref{Figure2}(c)). Here we see that the induced flow is respectively aligned or anti-aligned with the CW and CCW sound flow, leading to an effective Doppler shift. This implies that the acoustic mode with the same rotation direction as the induced chiral mass current will be frequency up-shifted, while that of opposite handedness is frequency downshifted. 
This can alternatively be viewed as arising from the fact that, while possessing an irrotational flow field, a sound wave can carry angular momentum, as more fluid advances under the peaks than recedes under the troughs \cite{forstner2019modelling}. The acoustic eigenmode where sound and chiral current angular momentum vectors are aligned therefore corresponds to a higher energy and frequency state than the one where these point in opposing directions.

We first estimate the magnitude $\Delta f$ of the splitting due to the dynamic coupling using a perturbative approach, following \cite{forstner2019modelling} 
(see also Methods).
We assume the distortion of the acoustic modeshape due to the chiral current to be small, so that the surface deflection profile $\eta\left(r,\theta,t\right)$ due to a  surface-sound wave of frequency $f=\Omega/2\pi$ confined to a circular domain of radius $R$ can be expressed in cylindrical coordinates as
\cite{baker2016theoretical,forstner2019modelling}:
\begin{equation}
\eta\left(r,\theta,t\right)=\eta_0\,J_m\left(\zeta \frac{r}{R}\right) e^{i\left(\,m\,\theta\pm \Omega\,t\right)}
\label{Eqcomplexeta}
\end{equation}
where $\eta_0$ is the wave amplitude, $J_m$ the Bessel function of the first kind, with $m$ the mode's azimuthal order, and $\zeta$ a parameter which depends on the mode radial order $n$ and the boundary conditions
 \cite{baker2016theoretical}. 
This approximation is justified if the ratio of the mass currents due to the chiral flow to that of the acoustic mode is small.
In our case, $\frac{g_{\mathrm{dyn}}}{g_s}=\frac{v_{\mathrm{dyn}}}{v_s} \propto \frac{\kappa_3}{R\,c_s}$, with $g_s$ the mass current associated with the surface wave (corresponding to the first term in Eq. \eqref{Eqmasscurrentmermin}), $c_s$ the speed of sound, and $\kappa_3=\frac{\pi\hbar}{M}$ the circulation quantum in $
^3$He. For the resonators considered here, this ratio does not exceed a few percent (see Methods).

We can directly identify the chiral current velocity from Eq. \eqref{eqinplanemasscurrent} as $\vec{v}_\mathrm{dyn}=\vec{g}_\mathrm{dyn}/\rho_0$.
The fluid-flow velocity associated with the surface waves, corresponding to the first term in Eq. \eqref{Eqmasscurrentmermin}, can be calculated from Eq. \eqref{Eqcomplexeta} to be $\vec{v}_s=\pm i c_s^2\vec{\nabla}\eta/(h_0\Omega)$, where '$+$' and '$-$' correspond to CW and CCW modes, respectively \cite{baker2016theoretical}.
The energy difference $\Delta E$ between the two sound eigenmodes arising from this overlap takes the form \cite{forstner2019modelling}:

\begin{equation}
\begin{aligned}
\Delta E_\mathrm{dyn}&=\frac{\rho_0 h_0}{2}\int\int\left(\vec{v}_\mathrm{s}+\vec{v}_{\mathrm{dyn}}\right)^2- \left(\vec{v}_\mathrm{s}-\vec{v}_{\mathrm{dyn}}\right)^2 \,r\,\mathrm{d}r\,\,\mathrm{d}\theta\\
&=2\rho_0 h_0\int\int \vec{v}_\mathrm{s}\cdot\vec{v}_{\mathrm{dyn}}  \,r\,\mathrm{d}r\,\mathrm{d}\theta.
\end{aligned}
\label{eq:deltaE}
\end{equation}
The resulting frequency splitting $\Delta f$ between the two counter-rotating sound modes is given by (see Methods 
for full derivation):
\begin{equation}
\Delta f_{\mathrm{dyn}} =\frac{\kappa_3\,m}{4\,\pi^2}\,\frac{\eta^2\left(R\right)}{\int_0^R \eta^2\left(r\right)\,r\,\mathrm{d}r}.
\label{Eqdeltaf}
\end{equation}
The splitting is independent of acoustic mode amplitude
and vanishes for fixed boundary conditions, for which $\eta\left(R\right)=0$.

The existence of an induced transverse flow in the hydrodynamic oscillation of a superfluid $^3$He-$A$ film had been noted~\cite{volovik1988analog}, and alluded to in the context of superfluid third sound resonators~\cite{schechter_observation_1998}. However, the splitting of sound modes scales inversely with resonator area 
(see Methods), and hence a measurable effect only becomes possible with the advent of the microscopic superfluid resonators \cite{harris_laser_2016,sachkou_coherent_2019,he_strong_2020} considered in this work.

\subsection*{Static chiral edge current}

The  static chiral edge current, which resides within a layer of a depth of the healing length $\xi=100$ nm from the resonator boundary in the case of an abruptly vanishing superfluid density is given by \cite{byun_measuring_2018,sauls_surface_2011} (see Methods)

\begin{equation}
\vec{g}_\mathrm{edge} = -\frac{\hbar}{4\,M} \vec{l}\times \vec{\nabla}\rho_\mathrm{edge} =\frac{\hbar\rho_0}{4\,M\,\xi}\hat{e}_{\theta},
\label{AppendixEqStaticChiralCurrent}
\end{equation}
where $\vec{\nabla}\rho_\mathrm{edge}$ quantifies the drop-off of the condensate density at the boundary. Here we can identify the superfluid velocity $\vec v_\mathrm{edge}=\vec{g}_\mathrm{edge}/\rho_0\approx 5$~cm/s, independently of the resonator geometry.

For a circular resonator with abruptly vanishing superfluid density (see e.g.~\cite{zhelev_-b_2017,levitin_phase_2013,Levitin_distortion_2013}), we find 
$
\Delta f_{\mathrm{edge}}
=-\Delta f_{\mathrm{dyn}}
$,
so that dynamic- and static contributions cancel out. An illustrative explanation is shown in Supplementary Figure 1.
On the other hand, if the film thickness
increases at the boundary, the contributions add constructively. 
The boundary can be engineered:
A reduced superfluid film thickness is associated with a stiffer van der Waals interaction, and such a constraint is typically associated with a fixed (or Dirichlet) boundary condition. A thickening of the film on the other hand is associated with a slower speed of sound, and a free (or Von Neumann) boundary condition for the film motion. In circular He-3 surface-sound resonators, both types of boundary conditions have been encountered~\cite{schechter_observation_1998},  the transition between the two regimes being associated with changes in film thickness~\cite{vorontsov_spectrum_2004}.
The latter configuration is realized which in our typical experimental designs, such as helium-coated microtoroids \cite{harris_laser_2016,sachkou_coherent_2019}
and silicon-on-insulator pillar designs \cite{Wasserman_cryogenic_2022}.

\subsection*{Experimental feasibility and perspectives}

 The first observation of $^3$He third sound employed a $R\sim2$~cm resonator, for which the magnitude of the splitting $\Delta f$ would have been in the hundreds of $\mu$Hz, and thus undetectable. Only recently,  advances in miniaturization and the fabrication of micrometer-sized third sound resonators~\cite{harris_laser_2016,sachkou_coherent_2019,he_strong_2020}, have put this observation within reach. 
 Indeed, for a circular resonator of 20 micron diameter and a film thickness of 100~nm, the magnitude of the dynamic splitting is appreciable:  47~Hz for the $(m=1;n=1;\Omega/(2\pi)=5.1$~kHz) mode and $555$~Hz for the $(m=6;n=1; \Omega/(2\pi)=70$~kHz) mode.
 This splitting would be detectable with sound quality factors on the order of 10$^2$ in both cases. For comparison, in superfluid $^4$He-resonators of the size and shape considered here, quality factors in excess of 10$^3$ are common \cite{harris_laser_2016, he_strong_2020}.

As a concrete example we 
propose silicon-on-insulator pillar designs, which we already fabricate for optomechanical experiments with surface-waves in helium-4 
, with standard pillar-height of 220~nm. 
These resonators have recently been shown to be able to confine superfluid surface-sound modes and allow for optomechanical readout \cite{Wasserman_cryogenic_2022}.
Sound waves arise from both vdW-and surface tension in comparable magnitudes (see Methods). 
The acoustic mismatch at the boundary leads to strong confinement for the higher order Bessel modes defined by the resonator geometry \cite{he_strong_2020,sachkou_coherent_2019}.
A range of readout mechanisms may be able to resolve the Majorana-induced splitting of sound waves, such as capacitive readout \cite{ellis_observation_1989} or torque sensing \cite{byun_measuring_2018}. Here, as a concetrete example, we focus on optical readout
via an optical whispering-gallery mode (WGM) confined on the perimeter of the pillar \cite{Wasserman_cryogenic_2022,Vilson_AllOptical_2004}.

Optomechanical experiments with $^3$He are inherently challenging because of its low superfluid transition temperature.
Heating issues can be minimized through techniques such as pulsed optical operation~\cite{Pulsed_Meenehan_2015}. 
Recently, through optimization of the superfluid fountain pressure interaction and high-precision optomechanical detection techniques, the ability to detect and strongly drive superfluid $^4$He third sound modes with optical powers down to the femtowatt range~\cite{sawadsky2022engineered}, corresponding to less than one average intracavity photon has been realized.
Mechanical modes are optically resolvable if the optomechanical cooperativity is larger than the inverse of the average thermal phonon number $C=4\bar n_\mathrm{cav}g_0^2/(\kappa\Gamma)>\bar n_\mathrm{th}^{-1}$, where $\bar n_\mathrm{cav}$ is the average intracavity photon number, $g_0$ is the single-photon optomechanical coupling rate, and $\kappa$ the optical decay rate \cite{Aspelmeyer_Review_2014}. A temperature of 1~mK yields $\bar n_\mathrm{th}=\hbar\Omega/(k_BT)=4.7\cdot10^3$ for the (m=1;n=1) mode and parameters considered above. With values from the experiment Ref. \cite{he_strong_2020}, we find $C=\bar n_\mathrm{cav}\cdot 2.5\cdot10^{-3}$, allowing the mechanical modes to be resolved with an average of less than one intracavity photon.

Of paramount importance in any experiment searching for Majorana modes, as demonstrated by recent discoveries of ambiguities \cite{Zhang_Retraction_2021} in claimed sightings of such modes, is the challenge of unambiguously distinguishing signatures of the Majorana mode from other effects \cite{Kayyalha_Absence_2020,Valentini_Majoranalike_2022}. Known effects that can produce a sound-mode splitting in superfluid helium resonators are geometric splitting and vortex-induced splitting \cite{sachkou_coherent_2019}. As the hydrodynamic equations describing momentum density for the $A$-and $B$-phases of superfluid $^3$He are identical except for the terms relating to the chiral Majorana current \cite{Vollhardt_Woelfle_1990},
these spurious sources of splitting can be calibrated out by transitioning the film between the $A$- and 
$B$-phases, by tuning temperature and film thickness \cite{levitin_phase_2013} or magnetic field. Hence,
any observed change in mode splitting between the $A$- and the $B$-phase can be attributed to chiral Majorana modes. 

Another potential limitation is the uniformity of the $\vec{l}$ vector.
For a thin film, the $\vec{l}$ vector can to first order only be aligned or anti-aligned with the surface normal. The twofold degeneracy in $\vec{l}$ can nevertheless lead to multiple domains in the film, whose opposite contributions to the sound coupling would cancel out. The splitting values quoted in the text are applicable to resonators covered by single domains, and constitute therefore an upper bound on the observable signal. The  use of microscale resonators, with dimensions much below the typical domain size~\cite{kasai_chiral_2018} increase here the likelihood of single domain coverage, and a specific cool-down procedure can be tailored for single-domain growth~\cite{byun_measuring_2018}.
The possibility of alignment of anisotropic superfluids by flow~\cite{de_gennes_alignment_1974} also offers the prospect of optomechanical control of the domains.

\section*{Discussion}
We have proposed an optomechanical method of measuring chiral Majorana fermions in quasi two-dimensional $^3$He-$A$ as the prototypical $p_x + i p_y$ superfluid. The sound can be considered as an oscillation of a deformed free surface in a superfluid film with the velocity along the film height gradient. A transverse surface current, orthogonal to the height gradient, is generated in a chiral superfluid with broken time reversal symmetry such as $^3$He-$A$. The transverse nature of this current couples two degenerate sound modes in orthogonal directions, and hence lifting their degeneracy. The resulting frequency splitting is perfectly suited for optomechanical detection based on the coupling between the sound and optical whispering gallery mode of micropillar optical resonator on which thin film of superfluid $^3$He-$A$ is deposited. We suggest a concrete realization of our proposal based on already existing silicon-on-insulator designs \cite{Wasserman_cryogenic_2022,Vilson_AllOptical_2004}. 

The splitting of the two modes is a direct consequence of the long-sought-after quantum Hall effect predicted in $^3$He-$A$. 
The Majorana-detection mechanism based on the surface waves relies on the coupling of deformation (strain) to a momentum current orthogonal to the strain direction. Therefore, the splitting can be viewed as a direct consequence of the quantum Hall viscosity which has not been experimentally verified despite various theoretical proposals. 
The deformation-induced current corresponds to the chiral Majorana current in a $p_x + ip_y$ superfluid.
Thus the proposed method offers a mean to detect the illusive chiral Majorana fermions.
 The flow of chiral Majorana fermions could be controlled if a particular surface deformation could be engineered by exciting and superposing different sound modes. 
This could be potentially useful in topological quantum computation based on braiding Majorana fermions. 
$^3$He-$A$'s intrinsic chiral nature may have other implications such as enabling topological acoustics~\cite{yang_topological_2015,souslov_topological_2017} and time-reversal symmetry breaking, without the need for any gain or drive of the medium~\cite{fleury_sound_2014}.

\section*{Methods}

\section*{Sound modes in a superfluid thin film}
\label{appendixsoundmodes}
For the superfluid resonators considered here, the relevant restoring forces are surface tension, van-der-Waals force and gravity. The respective speeds of sound are given by
\begin{equation}
c_\sigma=\frac{\zeta}{R}\sqrt{\frac{h_0\cdot\sigma_\mathrm{ST}}{\rho_0}}    \quad , \quad c_3=\sqrt{\frac{3\alpha_\mathrm{vdW}}{h_0^3}}
\quad \mathrm{and} \quad c_g=\sqrt{g_gh_0},
\end{equation}
with $\sigma_\mathrm{ST}=1.56\cdot 10^{-4}$~J/m$^2$ the surface tension of superfluid helium \cite{iino_surface_1985}
$\rho_0=82$~kg/m$^3$ the density of superfluid $^3$He \cite{schechter_third_nodate}, $\alpha_\mathrm{vdW}=3.5\cdot 10^{-24}$~m$^5$/s$^2$ the van der Waals coefficient of silicon, and $g_g=9.8$~m/s$^2$ the gravitational acceleration 
The total speed of sound is given by $c_\mathrm{s}=\sqrt{c_\sigma^2+c_3^2+c_g^2}$ and the sound frequency is
\begin{equation}
\nu=\frac{\Omega}{2\pi}=\frac{\zeta c_\mathrm{s}}{2\pi R}.   
\end{equation}
For the resonator dimensions and film thickness considered 
in this work (radius $R=10$~$\mu$m and film thickness $d=100$~nm), we find $f=2\pi\nu=5.1$~kHz for the (m=1;n=1) mode, and $f=70$~kHz for the (m=6;n=1) mode, with the gravitational contribution being negligible in this case \cite{mathematica_code}.

\section*{Validity of approximations}

\subsection*{Perturbative approach for the frequency shift}
\label{appendixsectionratio}

The perturbative approach used in the estimation of the frequency shift induced through both the dynamic and the static coupling between edge current and surface-sound modes is valid in the limit $v_\mathrm{s}\gg v_\mathrm{dyn}$. 
For a given film displacement amplitude $\eta_0$ we find
\begin{equation}
 v_{\mathrm{dyn}}\simeq \frac{\hbar \, \eta_0}{4\,M\, d \,R}\quad\mathrm{and}\quad
  v_\mathrm{s}\simeq \frac{c_\mathrm{s}^2 \, \eta_0}{\Omega\,d\,R}=\frac{c_\mathrm{s} \eta_0}{\zeta d}.
\end{equation}
The ratio between the two is given by $v_{\mathrm{dyn}}/v_\mathrm{s}=\frac{\kappa_3\,\zeta}{4\pi\,R\,c_\mathrm{s}}$ , which is on the order of one percent for the dimensions considered here, justifying the use of a perturbative approach to consider its effect on the sound waves.

\subsection*{In-plane and out-of-plane chiral currents}
\label{appendixinplaneoutofplane}
The chiral mass current is given by the second term of Eq. 
\eqref{Eqmasscurrentmermin}.
For a thin-film surface-wave resonator, it can be recast as
\begin{equation}
\vec{g}_\mathrm{dyn}=    \rho_0\frac{\hbar}{4\,M\,h_0}\vec\nabla\times (h_0+\eta) \,\vec{l}.
    \label{Eq_Appendix_gdyn}
\end{equation}
Simple vector calculus yields
\begin{equation}
\vec{\nabla}\times (h_0+\eta) \,\vec{l}=(h_0+\eta)\left(\vec{\nabla}\times\vec{l}\right)-\vec{l}\times\vec{\nabla}\eta.
\label{Eq_Appendix_rotVectorID}
\end{equation}
We consider, without loss of generality, that $\vec{l}$ is approximately aligned with $\hat{z}$.
As $\vec{l}$ remains normal to the film surface, we find

\begin{equation}
\vec{l}=\begin{pmatrix}
-\frac{\partial \eta}{\partial r}\\
-\frac{1}{r}\frac{\partial \eta}{\partial \theta}\\
1
\end{pmatrix} \quad \mathrm{and} \quad  \vec{\nabla}\times\vec{l}=\begin{pmatrix}
0\\
0\\
-\frac{\partial}{\partial r}\left(\frac{1}{r} \frac{\partial \eta}{\partial \theta} \right)+\frac{1}{r}\frac{\partial}{\partial \theta} \left( \frac{\partial \eta}{\partial r} \right)
\end{pmatrix}
\end{equation}
at the film surface.
Taking into account that $\vec{l}$ is aligned with the substrate at $z=0$, whereas it is aligned with the film surface at $z=h_0+\eta\approx h_0$, we find the anisotropy axis below the film's surface:
\begin{equation}
\hspace{-1cm}
\vec{l}=\begin{pmatrix}
-\frac{\partial \eta}{\partial r} \, \frac{z}{h_0}\\
-\frac{1}{r}\frac{\partial \eta}{\partial \theta}  \, \frac{z}{h_0}\\
1
\end{pmatrix} \quad \mathrm{and} \quad  \vec{\nabla}\times\vec{l}=\frac{1}{h_0}\begin{pmatrix}
\frac{\partial \eta}{\partial y}\\
-\frac{\partial \eta}{\partial x}\\
-\frac{\partial}{\partial r}\left(\frac{1}{r} \frac{\partial \eta}{\partial \theta} \right) z +\frac{1}{r}\frac{\partial}{\partial \theta} \left( \frac{\partial \eta}{\partial r} \right) z
\end{pmatrix}.
\label{Eq_Appendix_anisotroy2}
\end{equation}
Inserting Eq. \eqref{Eq_Appendix_anisotroy2} into Eq. \eqref{Eq_Appendix_rotVectorID} and neglecting terms proportional to $\eta^2$, we find 
\begin{equation}
\vec{\nabla}\times (h_0+\eta) \,\vec{l}\simeq -2\hat{z}\times\vec{\nabla}\eta.
\end{equation}
Inserting above equation into Eq. \eqref{Eq_Appendix_gdyn}, we obtain Eq. 
\eqref{eqinplanemasscurrent}.

The out-of-plane chiral current $\vec g_\mathrm{torque}$ is suppressed by a factor of $h_0/R$ (which is about 1\% for the resonators considered in this work)
compared to the in-plane current and will hence be neglected here.

\section*{Calculation of the dynamic coupling on a disk}
\label{appendixdynamiccoupling}
The energy difference $\Delta E$ between the two sound eigenmodes arising from a background flow is:
\begin{equation}
\begin{aligned}
&\Delta E\left(t\right)=\frac{1}{2} \, \rho_0\int_{\theta=0}^{2\pi}\int_{r=0}^R\int_{z=0}^{h_0+\eta\left(r,\,\theta,\,t\right)}\\& \left(\left|\left| \vec{v}_\mathrm{s}\left(\vec{r}, \, t \right)+\vec{v}_\mathrm{dyn}\left(\vec{r}\right)\right|\right|^2 -\left|\left| \vec{v}_\mathrm{s}\left(\vec{r}, \, t\right)-\vec{v}_\mathrm{dyn}\left(\vec{r}\right)\right|\right|^2 \right)r\,  \rm{d}r \,\rm{d}\theta \,  dz.
\end{aligned}
\label{EqdeltaEnumber1}
\end{equation}
Since $\vec{v}_\mathrm{s}$ and $\vec{v}_\mathrm{dyn}$ are independent of z, Eq.\eqref{EqdeltaEnumber1} becomes \cite{forstner2019modelling}:

\begin{equation}
\hspace{-0.5cm}\Delta E\left(t\right)=2\rho_0\int_{\theta=0}^{2\pi}\int_{r=0}^R   \vec{v}_\mathrm{s}\left(r,\theta, \, t\right)\cdot \vec{v}_\mathrm{dyn}\left(r,\theta\right) \left(h_0+ \eta\left(r, \theta, t \right)\right) r \, \rm{d}r  \, \rm{d}\theta
\label{EqdeltaEnumber2}
\end{equation}

Since $\eta\ll h_0$, we find:

\begin{equation}
\Delta E(t)
=2\rho_0h_0\int_{\theta=0}^{2\pi}\int_{r=0}^R  \vec{v}_\mathrm{s}\cdot\vec{v}_{\mathrm{dyn}}  \,r\,\mathrm{d}r\,\mathrm{d}\theta.
\end{equation}
Expressing the velocities as function of height perturbation (c.f. \cite{forstner2019modelling})
\begin{equation}
\vec{v}_\mathrm{s}=\pm \frac{i c_\mathrm{s}^2}{\Omega\,h_0}\vec{\nabla}\eta \quad \mathrm{and} \quad  \vec{v}_{\mathrm{dyn}}=\frac{\hbar}{4\,M\,h_0}\vec{\nabla}\times \eta \hat{z}
\label{eq:velocities}
\end{equation}
and using Eq. 
\eqref{Eqcomplexeta},
and performing straightforward vector calculus we obtain:
\begin{equation}
\Delta E(t)=\frac{\rho_0\,\hbar\,c_\mathrm{s}^2\,m}{2 M \,\Omega \,h_0} \int_0^R\int_0^{2\pi} \eta\left(r\right) \eta'(r) \\
\left( \cos^2\left(m\theta\pm \Omega\,t\right)+
\sin^2\left(m\theta\pm \Omega\,t\right)\right)  \,\mathrm{d}r\,\mathrm{d}\theta,
\end{equation}
where $\eta\left(r\right)=\eta_0 J_m\left(\zeta r/R\right)$ and $\eta'=\partial \eta/\partial r$ . Noting that $\Delta E$ is now independent of time, this simplifies to:
\begin{equation}
\Delta E=\frac{\pi\,\rho_0\,\hbar\,c_\mathrm{s}^2\,m}{M \,\Omega \,h_0} \int_0^R \eta\left(r\right) \eta'(r)  \,\mathrm{d}r.
\label{eqDeltaE}
\end{equation}
The energy $E$ stored in the sound mode is given for $m>0$ modes by \cite{baker2016theoretical}:
\begin{equation}
E=\frac{\pi \rho_0 c_\mathrm{s}^2}{2 h_0}\int_0^R \eta^2(r)\, r\, \mathrm{d}r.
\end{equation}
The fractional change in energy is thus:
\begin{equation}
\frac{\Delta E}{E}=\frac{2 \, \hbar\, m}{M \Omega}\frac{\int_0^R \eta\left(r\right) \eta'(r)  \,\mathrm{d}r}{\int_0^R \eta^2(r)\, r\, \mathrm{d}r}.
\end{equation}
For a harmonic oscillator, E is proportional to $\Omega^2$, and $\frac{\Delta E}{E}=2\frac{\Delta \Omega}{\Omega}$, hence the frequency splitting $\Delta f$ between the two counter-rotating sound modes is given by:
\begin{equation}
\Delta f =\frac{\Omega}{4\pi}\frac{\Delta E}{E}=\frac{\kappa_3\,m}{2\,\pi^2}\,\frac{\int_0^R \eta\left(r\right)\eta'\left(r\right)\,\mathrm{d}r}{\int_0^R \eta^2\left(r\right)\,r\,\mathrm{d}r}=\frac{\kappa_3\,m}{4\,\pi^2}\,\frac{\eta^2\left(R\right)}{\int_0^R \eta^2\left(r\right)\,r\,\mathrm{d}r}.
\label{EqdeltafSupp}
\end{equation}
Since $\eta(R)=0$ for fixed boundary conditions, this type of coupling only arises for free (or mixed) boundary conditions \cite{mathematica_code}. For a more general geometry, Finite-Element Modelling can be employed \cite{forstner2019modelling} 

The inverse scaling of the mode-splitting with resonator area can be understood as follows:
For a given wave amplitude, the magnitude of the surface gradient, and hence of the induced chiral flow velocity, is inversely proportional to R. The frequency shift $\Delta f_{\mathrm{dyn}}$ is proportional to $R^{-2}$. This can be seen from Eq. 
\eqref{Eqdeltaf} 
as, in average, $\eta^2(r)$ and $\eta^2(R)$ are independent of $R$, while $\int_0^R r dr \propto R^2$.

\section*{Calculation of the static coupling on a specular boundary}
\label{appendixstatic}
\label{appendixedgecurrent}
If $\rho$ drops from $\rho_0$ to 0 at the boundary within a length scale $\xi$, and $\vec{l}=\begin{pmatrix}
0\\
0\\
1
\end{pmatrix}$ is constant, then the chiral current density at the resonator boundary is 
\begin{equation}
\vec{g}_\textrm{edge} = -\frac{\hbar}{4\,M} \vec{l}\times \vec{\nabla}\rho_\mathrm{edge} =\frac{\hbar\rho_0}{4\,M\,\xi}\hat{e}_{\theta}
\label{AppendixEqStaticChiralCurrent}
\end{equation}
The angular momentum $L_z$ associated to this mass current density $\vec{g}_\textrm{edge}$ is, with $R>>\xi$:
\begin{equation}
L_z=m\,v\,R=g_\mathrm{edge}\,(2\pi\,R)\,\xi\,R=\frac{\hbar}{2}\frac{\pi\,R^2 \,\rho_0}{M}=\frac{\hbar}{2}\,N,
\end{equation}
where N is the number of particles in the fluid.

From Eq. \eqref{AppendixEqStaticChiralCurrent}, we can identify the velocity of the edge current

\begin{equation}
\vec{v}_\mathrm{edge}\simeq\frac{\hbar}{4\,M\,\xi} \hat{e}_{\theta}.
\end{equation}
For the case of an abrupt and specular boundary, $\xi$ is given by the healing length $\xi\sim 100$~nm. In this case, $v_\mathrm{edge}$ is on the order of 5~cm/s. The presence of the edge current leads to a background flow, which will lead to an energy difference $\Delta E$ between CW and CCW modes. We follow Ref. \cite{forstner2019modelling} in calculating the resulting frequency splitting:

\begin{equation}
\Delta E=2\rho_0\int_{\theta=0}^{2\pi}\int_{r=0}^R (h_0+\eta)\, \vec{v}_s\cdot\vec{v}_{\mathrm{edge}}\,\, r \,\mathrm{d}r\,\mathrm{d}\theta.
\end{equation}
Due to symmetry, $\iint  \vec{v}_s \cdot \vec{v}_\mathrm{edge}=0$, and hence \begin{equation}
\Delta E=2\rho_0\int_{\theta=0}^{2\pi}\int_{r=0}^R \eta(r)\, \vec{v}_s\cdot\vec{v}_{\mathrm{edge}}\,\, r \,\mathrm{d}r\,\mathrm{d}\theta.
\label{eq:static_coupling_velocity_product}
\end{equation}
The energy difference $\left<\Delta E\right>$ averaged over an oscillation period T is given by:
\begin{equation}
\begin{aligned}
&\left<\Delta E\right>= \frac{1}{T} \int_0^T \Delta E\left( t \right)  t\\&= 2\,\rho_0\int_{\theta=0}^{2\pi}\int_{r=0}^R r  \left(\vec{v}_\mathrm{edge}\cdot\hat{e}_r \, \frac{1}{T} \int_0^T \vec{v}_s\cdot\hat{e}_r \, \eta \, \mathrm{d} t + \vec{v}_\mathrm{edge}\cdot\hat{e}_\theta \, \frac{1}{T} \int_0^T \vec{v}_s\cdot\hat{e}_\theta \, \eta \, \mathrm{d} t \right)  \,\mathrm{d}r\,\mathrm{d}\theta
\end{aligned}
\label{EqDeltaEStatic1}
\end{equation}
From Eqs. 
\eqref{Eqcomplexeta} 
\& \eqref{eq:velocities} we see that $\vec{v}_s\cdot\hat{e}_r$ and $\eta$ are out-of-phase and $\vec{v}_s\cdot\hat{e}_\theta$ and $\eta$ are in-phase. Hence, the first part of the integral in Eq. \eqref{EqDeltaEStatic1} is zero and the remaining integral becomes $\frac{1}{2}\vert\vec{v}_s\cdot\hat{e}_\theta\vert\vert\eta\vert$.

Performing the integral and inserting Eqs. 
\eqref{Eqcomplexeta} 
\& \eqref{eq:velocities}, we find the same form as in Eq. \eqref{eqDeltaE}, and hence the same magnitude of spitting as for the dynamic case:
\begin{equation}
\Delta f_{\mathrm{edge}} =\frac{\kappa_3\,m}{4\,\pi^2}\,\frac{\eta^2(R)}{\int_0^R \eta^2\left(r\right)\,r\,\mathrm{d}r}.
\end{equation}
It is important to note here that $\Delta f_{\mathrm{edge}}$ is calculated with respect to CW and CCW rotating modes, whereas $\Delta f$ in the previous section was computed in the standing mode basis. As discussed in the main text, for the simple case of vanishing superfluid density at the boundary, the two contibutions have opposite signs and cancel each other.

\section*{Data availability}
The data that support the findings of this study are available within the paper and its Supplementary Information files. The Mathematica-code used in this work is available on Zenodo \cite{mathematica_code}.

\section*{Author contributions}
C.G.B., H.C., and W.P.B. provided overall leadership for the project. S.F., H.C., and C.G.B. conceptualized the idea, which was further developed by all co-authors. S.F. and C.G.B. performed analytical and numerical calculations. All co-authors contributed in discussions and in the development of the manuscript, which was drafted by S.F., C.G.B., and H.C..    

\section*{Acknowledgements}
This work was funded by the US Army Research Office through grant number W911NF17-1-0310 and the Australian Research Council Centre of Excellence for Engineered Quantum Systems (EQUS, project number CE170100009). C.G.B and G.I.H acknowledge the Australian Research Council Fellowship DE190100318 and DE210100848, respectively. 
S.F acknowledges the European Union (MSCA, MechQSim, 101105814) Marie Curie Fellowship. H.C acknowledges support from Samsung Science \& Technology Foundation SSTF-BA1601-08 and NRF 2022R1A2C2010750. The authors acknowledge helpful discussions with Utso Bhattacharya and Gian-Marco Schn\"{u}ringer. 
\section*{Additional information}

{\bf Competing interests:} The Authors declare no competing interests.

\newpage
\bibliographystyle{ieeetr}
\bibliography{bibliography_tidied}

\pagebreak
\section*{Supplementary Note 1. Transformations for first sound} 
 
\label{appendixfirstsound}
The method for detection of chiral current proposed in this work could also be applied to density waves in a three-dimensional superfluid $^3$He resonator. Using current technology, this approach appears significantly less promising than using surface waves for the following reasons: Firstly, a resonator without boundaries, as proposed in the main test, can not be realized, so that the unpredictable effect of the edge current has to be taken in to account when interpreting experimental results. Secondly, our preliminary calculations show that the expected frequency shift to sound modes in the bulk would be several orders of magnitude smaller than for surface waves, and hence very hard to detect. Nevertheless, it might be useful to consider this effect in future experiments, including long-term goals such as braiding of Majorana fermions or topological acoustics.

\begin{center}
\begin{table}[htp!]
\begin{tabular}{ c  c} 
 \hline
 Surface acoustics\quad &\quad Bulk acoustics    \\ [0.5ex] 
 \hline\hline
 film height perturbation & density perturbation  \\ 
 $\eta(\vec r, t)$ & $\delta\rho(\vec r, t)$\\
 \hline
 unperturbed film height & static density  \\ 
 $h_0$ & $\rho_0$\\
 \hline
 \textit{dynamic} chiral current(1-D) & \textit{dynamic} chiral current(2-D)  \\ 
 $\vec{g}_\mathrm{dyn,1D}=-\rho_0\frac{\hbar}{2\,M\,h_0}\hat{z}\times \vec{\nabla}\eta+g_\mathrm{torque}$ \quad & \quad $\vec{g}_\mathrm{dyn,2D}=-\frac{\hbar}{2\,M}\hat{l}\times \vec{\nabla}\rho+g_\mathrm{torque}$
 \\
 \hline
 \textit{static} chiral current(1-D) & \textit{static} chiral current(2-D)  \\ 
 $\vec{g}_\mathrm{edge} =\frac{\hbar\rho_0}{4\,M\,\xi}\hat{e}_{\theta}$ & $\vec{g}_\mathrm{surface} =\frac{\hbar\rho_0}{4\,M\,\xi}\hat{l}\times \hat{n}$ 
 \\
 \hline
 surface speed of sound & bulk speed of sound  \\
  $c_\mathrm{s}=\sqrt{c_\sigma^2+c_3^2}$ (see Appendix \ref{appendixsoundmodes}) & $c_\mathrm{bulk}=\sqrt{K_1/\rho_0}$\\
 \hline
\end{tabular}
\caption{Transformations for chiral current dynamics between surface waves and bulk density waves. $\hat{n}$ is the surface normal of the three-dimensional resonator, and $K_1$ the bulk modulus of first sound in superfluid $^3$He.}
\label{tableBulkSurface}
\end{table}
\end{center}
\clearpage
\pagebreak

\section*{Supplementary Note 2. Illustration of the cancellation of dynamic- and edge current in a disk resonator with abrupt, specular boundary}

\begin{figure}[b!]
\centering
    \includegraphics{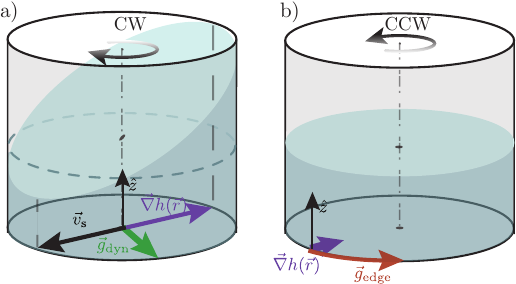}
    \caption{(a) Illustration of the directionality of dynamic chiral current. $\vec{\nabla}h(\vec r)=\vec{\nabla}\eta(\vec r)$ (purple arrow) points toward the peak of the acoustic wave. The direction of the dynamical chiral current (green arrow) is given by $-\hat{z}\times \vec{\nabla}h(\vec r)$, so that it points toward the right-hand side of the peak, increasing the energy of the clockwise (CW) rotating acoustic mode (curved black arrow). The trough and peak of the acoustic mode are indicated by dashed vertical lines, and the sound velocity is indicated by a black arrow. All arrows are shown on the bottom of the superfluid resonator for ease of illustration, although the fluid flow is distributed through the whole film.  
    (b) Illustration of the directionality of the static edge current. $\vec{\nabla}h(\vec r)=-\hat{e}_r h_0\delta(r-R)$ (purple arrow) points toward the inside of the resonator at its boundary. The direction of the edge current (curved red arrow) is given by $-\hat{z}\times \vec{\nabla}h(\vec r)$ and points in the counterclockwise (CCW) direction, hence increasing the energy of the CCW acoustic mode.  By analogy, if the film thickness increases at the resonator boundary (so that $\vec{\nabla}h(\vec r)$ points outward, not shown in figure),the static part of the chiral current increases the energy of the CW mode. While (a) and (b) refer to the same resonator, in (b) the deformation due to the sound mode is not shown to clarify that it does not cause any static current.
    }
    \label{Suppfig:statdyncancellation}
\end{figure}

\end{document}